# Experimental Probing of Photonic Density of States in Hyperbolic Metamaterial

M. A. Noginov[1*], H. Li[1], D. Dryden[2], G. Nataraj[2], Yu. A. Barnakov[1], G. Zhu[1], M. Mayy[1], Z. Jacob[3], E. E. Narimanov[3]

[1] *Center for Materials Research, Norfolk State University, Norfolk, VA 23504*
[*] *mnoginov@nsu.edu*
[2] *Summer Research Program, Center for Materials Research, Norfolk State University, Norfolk, VA 23504*
[3] *Electrical Engineering, Purdue University, West Lafayette, IN 47907*

**Abstract:** In the metamaterial with hyperbolic dispersion (an array of silver nanowires in alumina membrane) we have observed six-fold reduction of the emission life-time of dye deposited onto the metamaterial's surface. This serves as the evidence of the earlier predicted high density of photonic states in hyperbolic metamaterials.

*Metamaterials* are engineered composite materials containing sub-wavelength inclusions (meta-"atoms") that often have custom-tailored shapes, sizes, mutual arrangements, and orientations. Near-fantastic theoretical predictions and experimental demonstrations include but are not limited to negative index of refraction, focusing and imaging with infinitely high resolution, and invisibility cloaking. Metamaterials with hyperbolic dispersion [1-3], in which different elements of the dielectric tensor have different signs, enable the phenomena of negative refraction and hyperlens [4,5]. Hyperbolic materials are predicted to have broad-band singularity of photonic density of states, which causes enhanced and highly directional spontaneous emission and enables a variety of devices with new functionalities, including a single-photon gun [6]. In this work, we probed the photonic density of states in hyperbolic metamaterials experimentally and found the experimental results to be in a qualitative agreement with the theoretical predictions [6].

Alumina membranes with the dimensions 1cm x 1cm x 51μm, acquired from Synkera Inc., had 35 nm channels (voids) extending through the whole thickness of the membrane perpendicular to its surface. The surface filling factor of the voids was ~15%. The membranes were filled with silver, following the method described in Ref. [3]. This synthesis technique results in a metamaterial with hyperbolic dispersion (negative electric permittivity in the direction perpendicular to the membrane's surface, $\varepsilon_\perp$, and positive electric permittivity in the direction parallel to the membrane, $\varepsilon_\parallel$) in the near-infrared range of the spectrum [3]. The hyperbolic dispersion of the samples fabricated in this work has been confirmed for the substantial part of the emission band of dye, discussed below, in the measurements of reflectance versus incidence angle carried in both p and s polarizations [3].

In experiments, IR 140 laser dye-doped PMMA polymeric films have been deposited on the top of the silver-filled membranes. The emission band of the IR140 dye has its maximum at 892 nm and the full width at half maximum (FWHM) equal to 37 nm. The absorption band has the peak at 820 nm and FWHM equal to 90 nm. Control samples included dye-doped polymeric films deposited onto pure glass, neutral density glass filter with the absorption coefficient ~50 cm$^{-1}$, pure (unfilled) membrane, and silver and gold thin films on glass. The thickness of the dye-doped polymer film on the mechanically robust control samples (evaluated based on the DekTak-6 profilometer measurements performed on control samples) was approximately equal to 130 nm. The samples were excited at $\lambda$=800 nm with 100 fs pulses of the mode-locked Ti-sapphire laser (Mira-900) and the emission was detected with the C5680 streak camera. The emission was separated from pumping by the 810 nm long-pass filter.

The emission kinetics in the dye-doped film deposited onto unfilled membrane was nearly single-exponential, with the decay-time equal to 760 ps, Fig. 1. The emission kinetics in some of the other control samples, *e.g.* gold film, had a peak overlapping with the pumping pulse and their exponential decay kinetics were shortened to ~700 ps, Fig. 1. The short peak in the beginning of the emission kinetics (also observed in substrates without dye-doped films) is a combination of the pumping light leaking through the filter as well as luminescence and nonlinear optical response of the substrate to 100 fs laser pulses and is not of interest to the present study. In the film deposited onto the silver-filled membrane, the emission decay-time of dye (averaged over several measurements taken in different parts of the sample), was as short as ~125 ps, Fig. 1, in a qualitative agreement with the theoretical predictions [6]. The shortening of the emission decay-time could not be due to non-radiative luminescence quenching on silver inclusions, since no quenching of comparable magnitude was observed in the samples deposited onto metallic films. Since the emission kinetics shortening is explained by the enhancement of emission irradiated *inside* the metamaterial sample, the emission intensity, detected from the same side of the sample from which it is illuminated, is expected to get reduced. This prediction is in an agreement with the experimental observation, Fig. 1.

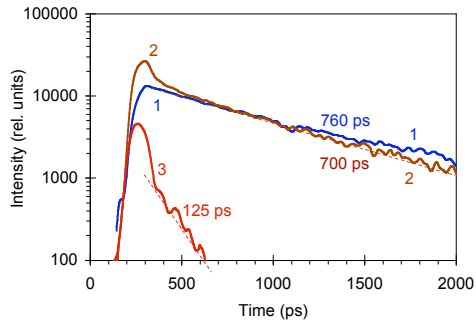

Fig. 1. Emission kinetics in the IR140/PMMA films deposited on the (1) top of pure alumina membrane, (2) gold film on glass and (3) silver-filled alumina membrane.

The experimentally observed shortening of the emission kinetics was stronger than that predicted theoretically [6]. The difference can be explained by a variety of reasons, including underestimation of the negative value of $\varepsilon_\perp$ and overestimation of the thickness of the dye-doped polymeric film.

The experimental evidence of the anomalously high photonic density of states in hyperbolic metamaterials paves the road to many exciting applications including a single photon gun needed for quantum optics and information technology [6].

The work was supported by the NSF PREM grant # DMR 0611430, NSF NCN grant # EEC-0228390, AFOSR grant # FA9550-09-1-0456, and ARO-MURI award 50342-PH-MUR. The authors cordially thank C. E. Bonner for the assistance with the kinetics experiments.